\documentclass[preprint]{elsarticle}
\usepackage{amsmath}
\usepackage{amssymb}
\usepackage{epsfig}
\usepackage{epstopdf}
\usepackage[latin5]{inputenc}
\usepackage{subfigure}
\usepackage{lineno,hyperref}

\modulolinenumbers[5]

\usepackage{color}

\numberwithin{equation}{section}

\begin{document}

\title{Dispersive shock waves in Three Dimensional Benjamin-Ono equation}

\author{Ali Demirci}
\ead{demircial@itu.edu.tr}

%\cortext[cor1]{Corresponding author}
\address{Department of Mathematics, Faculty of Science and Letters, Istanbul Technical University, Istanbul 34469, Turkey}

\begin{abstract}
Dispersive shock waves (DSWs) in the three dimensional Benjamin- Ono (3DBO) equation is studied with step-like initial condition along a paraboloid front. By using a similarity reduction, problem of studying DSWs in three space one time (3+1) dimensions reduces to finding DSW solution of a (1+1) dimensional equation. By using a special ansatz, the 3DBO equation exactly reduces to the spherical Benjamin-Ono (sBO) equation. Whitham modulation equations are derived which describes DSW evolution in the sBO equation by using a perturbation method and these equations are written in terms of appropriate Riemmann type variables to obtain the sBO- Whitham system. DSW solution which obtained from the numerical solutions of the Whitham system and the direct numerical solution of the sBO equation are compared. In this comparison, a good agreement is found between these solutions. Also, some physical qualitative results about DSWs in sBO equation are presented. It is concluded that DSW solutions in the reduced sBO equation provide some information about DSW behaviour along the paraboloid fronts in the 3DBO equation.
\end{abstract}


\date{\today}

\maketitle

\section{Introduction}
Dispersive shock waves (DSWs) have been studied extensively in recent years, because of their applications/observations in water wave theory \cite{Ligh78}, \cite{Smy88} (termed as undular bore), plasma physics \cite{Tay70} (termed as collisionless shock wave), nonlinear optics \cite{Wan07, Con09, Fat14} and Bose- Einstein condensates (BEC) \cite{Hoe06, Hoe08} ,etc. A canonical problem for existence of DSWs is the Korteweg-de Vries (KDV) equation with small dispersion and initial conditions corresponding to a unit step (Heaviside) function. Gurevich and Pitaevskii \cite{Gru74} investigated this problem by employing an averaging method which was pioneered by Whitham \cite{Whi65}. Over the years there have been numerous important analytical and numerical studies about DSWs that employ Whitham methods to some partial differential equations (PDEs) such as the KdV equation \cite{Lax83} and the mKdV equation \cite{Dris76}, the KdV-Burgers \cite{Gru87} equation, the Benjamin- Ono (BO) equation \cite{Mat98} and the BO- Burgers equation \cite{Mat07}, and the Gardner equation \cite{Kamc12}. For an extensive review of most of these works we refer to the paper \cite{Hoe16} and references therein. All of these studies were restricted to (1+1) dimensional PDEs and much less information had been known about DSWs in multidimensional PDEs until recently.

In the last decade, DSWs in two space one time (2+1) dimensional systems have been subject of many studies. In \cite{EL09a,Hoe09a}, 2+1 dimensional NLS type equations and associated DSW solutions were analyzed by reducing them to associated 1+1 dimensional systems. Also, DSWs in the Kadomtsev-Petviashvili (KP) \cite{KP70} and the two dimensional Benjamin-Ono (2DBO) \cite{Abl80} equations were considered using step like initial data along a parabolic front \cite{Dem16}. In \cite{Dem16}, by using a similarity variable (1+1) dimensional cylindrical reductions of these equations and their associated DSW solutions were investigated. Associated Whitham systems of these cylindrical equations were derived and studied the DSW solutions with step like initial data in terms of relevant Riemann type variables. We note that the method used in \cite{Dem16} works only under the special choice of parabolic front. Later, a generalization of Whitham theory for DSWs in the KP- type equations with any initial conditions was developed \cite{Abl17, Abl18}. The main result of these works, the $5x5$ system of (2+1)- dimensional hydrodynamic- type equations \cite{El13a} was derived which describes the slow modulations of the periodic solutions of the corresponding KP- type equation. The method presented in \cite{Abl17, Abl18} can be applied to DSW investigation in integrable and non- integrable (2+1) dimensional PDEs.

In the present study, we consider DSWs in three dimensional Benjamin- Ono (3DBO)  equation (see Eq. \ref{kpbo}) by employing a similarity reduction and thus reducing this equation to the (1+1) dimensional spherical Benjamin- Ono equation (see Eq. \ref{sBO}). The main motivation of this paper is to understand DSW behavior in the 3DBO Eq. when there is step like data along paraboloid front in the initial data. The method presented in \cite{Dem16} is used to study this problem. We analyze the sBO equation via Whitham theory and derive Whitham modulation equations. Then, these equations are transformed into simpler form by using appropriate Riemann type variables. These transformed Whitham equations in Riemann variables are not in diagonal hydrodynamic form \cite{El13a}. This diagonal property is related to integrability property of the main PDE system and neither 3DBO equation nor its reduction, the sBO equation is known to be integrable.

Since the Whitham system is non-diagonal, it is solved numerically and obtain the DSW solution of the sBO equation with these numerical solutions. We then compare these (1+1) dimensional DSW structure to direct numerical simulations of the sBO equation. After fixing parameters our comparisons between these solutions exhibit very good results. Also, the solution of the 1+1 dimensional spherical equation is shifted by a value which is determined by the solution of the Front Shape equation (see Eq. \ref{peq}) for the desired circular cross section and time $t$. By this shifting procedure, the evolution of DSWs on any cross section of the initial paraboloid in the 3DBO equation can be constructed for any time value $t$.

To our knowledge this is the first time the investigation of DSWs in a 3+1 dimensional system is analyzed in detail. We must note that the 3DBO equation has not derived exactly in any research area yet. However, the 3DBO equation can be considered just a generalization of the 2DBO equation \cite{Abl80} and investigations of the DSWs solutions in this equation may help to generalize Whitham modulation theory to 3+1 dimensional PDEs to study 3+1 dimensional DSWs.

This paper is organized follows. In Section 2 a similarity reduction is used to exactly transform the 3DBO equation to the sBO equation along a paraboloid front. In Section 3 we employ perturbation theory \cite{Luk66} to find hydrodynamic type equations associated with Whitham theory for the BO and sBO equations. By employing appropriate Riemann type variables, we then transform the Whitham modulation equations. Since this transformed Whitham system is not in diagonal form, we solve it numerically and reconstruct the DSW solutions of BO and sBO. We then compare the constructed solutions with direct numerical solutions of BO and sBO. Apart from an unimportant phase, these compared solutions are in very good agreement. We also report that solution of Whitham system for sBO indicate a small discontinuity. This discontinuity would be resolved by taking into account higher order terms (for details see \cite{Abl70}), but it is out of scope of this study. Then we devote to concluding remarks and future studies in the last section.

\section{Reduction of 3DBO equation  to sBO equation}
In this section, we study DSW propagation associated with the $(3+1)$ dimensional equation
\begin{equation}
\label{kpbo}
\left(u_t+uu_x+\epsilon \mathcal{H}\left(u_{xx}\right)\right)_{x}+ \lambda \left( u_{yy}+u_{zz}\right)=0
\end{equation}
where $\mathcal{H}u(x)$ denotes the Hilbert transform:

\begin{equation}
\label{Hilbert}
\mathcal{H}u(x)= \frac{1}{\pi}\mathcal{P}\int_{-\infty}^{\infty} \frac{u(x')}{x'-x}dx'
\end{equation}
and $\mathcal{P}$ denotes the Cauchy principal value. We refer to  Eq.~(\ref{kpbo}) as the 3DBO (Three Dimensional Benjamin-Ono) equation; it is a  three-dimensional extension of the classical BO equation. Two dimensional version of this equation describes weakly nonlinear long internal waves in fluids of great depth \cite{Abl80}. Here $\epsilon, \lambda$ are constant. When $|\epsilon| \ll1$ the system has weak dispersion and according to sign of $\lambda$. According to the sign of $\lambda$, Eq.~(\ref{kpbo})  describes dynamics of strong surface tension of the water for ($\lambda=-1$) and weak surface tension for ($\lambda=1$) , respectively.

Also, we are interested in a class of  initial conditions for Eq.~(\ref{kpbo}) which is the three dimensional extension of the Riemann-type initial condition
\begin{equation}
\label{ric}
u(\eta,0) = \left\{
\begin{array}{lr}
 1, & \eta <0;\\
 0, &  \eta \geq0,
 \end{array}
 \right.
\end{equation}
where $\eta=x+\frac{1}{2}P(y,z,0)$.

The goal in this paper is to improve understanding of DSWs in a (3+1) dimensional nonlinear partial differential equation (PDE) with a special type of initial condition. For this purpose, we consider a reduction of the Eq.~(\ref{kpbo}) for the special choice of a paraboloid initial front
\begin{equation}\label{eq:Pcy2}
P(y,z,0)=\tilde{c}\left(y^2+z^2\right),
\end{equation}
where $\tilde{c}$ is a real constant.

We assume that solutions of Eq.~(\ref{kpbo}) satisfy the following ansatz:
\begin{equation}
\label{anz}
u=f(x+P(y,z,t)/2,t);
\end{equation}
the front is then described by $x+P(y,z,t)/2=$ constant. We substitute the  ansatz (\ref{anz}) into Eq.~(\ref{kpbo}) and find
\begin{equation}
\label{rel1}
\left(\frac{1}{2}P_{t}f_{\eta}+f_t+ff_{\eta}+\epsilon \mathcal{H}\left(f_{\eta\eta}\right)\right)_{\eta}+\lambda\left(\frac{1}{4}\left(\left(P_{y}\right)^2+\left(P_{z}\right)^2\right)f_{\eta\eta}+\frac{1}{2}\left(P_{yy}+P_{zz}\right)f_{\eta }\right)=0
\end{equation}
where $\eta=x+P(y,z,t)/2$. In the Eq.~(\ref{kpbo}) $u$ satisfies the following boundary conditions at the infinities:
\begin{equation}
\label{bc2}
u\rightarrow R(t)\,\,\,\, \textrm{as} \,\,\, \eta\rightarrow-\infty\,\,\,\,\, \textrm{and} \,\,\,\,\,\,u\rightarrow0\,\,\,\, \textrm{as} \,\,\,\,\,\eta\rightarrow\infty.
\end{equation}
The function $R(t)$ is chosen appropriately; the forms of  $R(t)$ are determined later in this section.

Using these boundary conditions and assuming that $P_{yy}$ and $P_{zz}$ is independent of $y$ and $z$, the coefficient of the term $f_{\eta}$ depends only $t$ in Eq.~(\ref{rel1}). If we also assume that the sum of the coefficients of the term $f_{\eta\eta}$ vanishes in Eq.~(\ref{rel1}), then we obtain the following system of equations
\begin{subequations}
\label{eqs}
\begin{equation}
\label{peq}
P_t+\frac{\lambda}{2}\left(\left(P_y\right)^2+\left(P_z\right)^2\right)=0,\,\,\,(front\, shape\, equation:\, FS)
\end{equation}
\begin{equation}
\label{feq}
f_{t}+ff_{\eta}+\frac{\lambda}{2}\left(P_{yy}+P_{zz}\right)f+\epsilon \mathcal{H}\left(f_{\eta\eta}\right)=0.
\end{equation}
\end{subequations}
The solution of the initial value problem (IVP) for the FS equation (\ref{peq}) and (\ref{eq:Pcy2}) is found as
\begin{equation}
\label{vs1}
P(y,z,t)=\frac{\tilde{c}\left(y^2+z^2\right)}{1+2\tilde{c} \lambda t}.
\end{equation}
This solution gives a consistent result about assumption on $P_{yy}$ and $P_{zz}$. Both these terms are independent of $y$ and $z$ for all time.

The level surfaces of the solution $\eta=x+\frac{P(y,z,t)}{2} = \eta_0$,  $\eta_0$ constant, are along the following surfaces in the $(x,y,z)$-space at fixed time $t$:
\begin{equation}
\label{center}
x= -\frac{\tilde{c}\left(y^2+z^2\right)}{2(1+2\lambda \tilde{c}t)} +\eta_0= C(t)\left(y^2+z^2\right)+\eta_0, ~~C(t)=- \frac{\tilde{c}}{2(1+2\lambda \tilde{c}t)}.
\end{equation}
For the case $\tilde{c} \lambda >0$,  $C(t)$ indicates the paraboloid front `flattens' when $t$ increases. However, the function $C(t)$ blows up  at a critical time $t_c=-1/(2 \lambda \tilde{c})$. In this study, we will consider $\tilde{c} \lambda>0$ and only be concerned with $t>0$. This blow up phenomena is out of scope of this study and it should be investigated for a prospective study in details.

Then, we substitute the front solution (\ref{vs1}) into Eq.~(\ref{feq}) and obtain the following the spherical Benjamin-Ono (sBO) equation:
\begin{equation}
\label{sBO}
f_{t}+ff_{\eta}+\frac{2\lambda \tilde{c}}{1+2\lambda \tilde{c}t}f+\epsilon\mathcal{H}\left(f_{\eta\eta}\right)=0.
\end{equation}
In summary, it has been shown that for the paraboloid front initial condition (\ref{eq:Pcy2}), by using the ansatz (\ref{anz}), the (3+1) dimensional PDE (\ref{kpbo}) is reduced to a (1+1) dimensional PDE (\ref{sBO}) with variable coefficients. Similar reduced equation can be obtained by applying to the similarity reduction approaches of PDEs \cite{Gun18}. to Eq. (\ref{kpbo}).

We will denote $t_0=\frac{1}{2\lambda \tilde{c}}$, then the term $\frac{2\lambda \tilde{c}}{1+2 \lambda \tilde{c}t}$ is written as $\frac{1}{(t+t_0)}$. Also, we will consider only $\lambda=1$; the other sign can be obtained by changing $\tilde{c}$ to $-\tilde{c}$.

For identifying the boundary conditions associated with Eq.~(\ref{sBO}) at infinity, we omit $\eta$ dependent terms and then solve the remaining ODE with the associated initial condition (\ref{ric}). The solution of this ODE with the initial condition $R(0)=1$ determines the function $R(t)$ in the boundary condition (\ref{bc2}) as
\begin{equation}
\label{rr}
R(t)=\frac{t_0}{t+t_0}.
\end{equation}

In the next section, we examine the DSW solutions of Eq. (\ref{sBO}) with (non-increasing) initial data  such as Eq.~(\ref{ric}).

\section{Dispersive shock waves in BO and sBO equations}

\subsection{Whitham modulation equations for BO/ sBO equations}

In this section we will investigate the DSW solutions of the BO and sBO equations by using the Whitham modulation theory. We will compare Whitham theory for both these equations at leading order and direct numerical simulations.

In the use of Whitham modulation theory we will find three conservation laws. Then, we transform these three conservation laws into a system of quasilinear first order PDEs (Whitham system) by using convenient Riemann type variables. This approach was first introduced by Whitham for the KdV equation \cite{Whi65}. For the integrable equations such as KdV eq. this system  can be diagonalized and solved exactly. But neither 3DBO eq. nor its reduction sBO eq. is known to be integrable eq. according to our knowledge. Numerically we show that the associated Whitham system has solution which demonstrate the DSW structure of sBO, unlike BO, decays in time.

We will use a method of multiple scales which was used by Luke \cite{Luk66} in the investigation of Whitham type systems associated with a nonlinear Klein-Gordon equation. By following the approach of Luke, we obtain a periodic wave solution of the leading order equation for the sBO eq. This periodic solutio has three independent free parameters which are slowly varying. The leading order problem introduces the rapidly varying phase which requires a compatibility condition which is often termed conservation of waves.  The next order problem in the perturbation method has two secularity conditions; these together with conservation of waves provide three necessary Whitham modulation equations.

For this purpose, first we assume $f=f(\theta, \eta,t;\epsilon)$. Here $\theta$ is rapidly varying and defined from

\begin{equation}
\label{phase}
\theta_{\eta}=\frac{k(\eta,t)}{\epsilon},\,\,\,\,\,\,\theta_{t}=-\frac{\omega(\eta,t)}{\epsilon}=-\frac{kV}{\epsilon}
\end{equation}
where $k$, $\omega$ and $V$ are the wave number, frequency and phase velocity, respectively. This definition provides us the compatibility condition $\left(\theta_{\eta}\right)_t=\left(\theta_t\right)_{\eta}$ (conservation of waves) as
\begin{equation}
\label{comp}
k_t+\left(kV\right)_{\eta}=0.
\end{equation}
This is the first necessary eq. for modulation eqs.

We transform Eq. ~(\ref{sBO}) with these slow and rapid varibles to the following eq.:
\begin{equation}
\label{orderbo}
\begin{aligned}
&\frac{1}{\epsilon}\left[-\omega\frac{\partial f}{\partial \theta}+k f \frac{\partial f}{\partial \theta}+k^2\mathcal{H}\left(\frac{\partial^{2} f}{\partial \theta^{2}}\right)\right]\\
&+\left[\frac{\partial f}{\partial t}+f\frac{\partial f}{\partial \eta}+\mathcal{H}\left(k_{\eta}\frac{\partial f}{\partial \theta}+2k\frac{\partial^{2} f}{\partial \theta \partial \eta}\right)+\frac{1}{t+t_0}f\right]+\epsilon \mathcal{H}\left(\frac{\partial^2 f}{\partial \eta^{2}}\right)=0.
\end{aligned}
\end{equation}
As usual in the method of multiple scales, then we expand $f$ in powers of $\epsilon$  as

\begin{equation}
\label{exp2}
f\left(\theta,\eta,t\right)=f_0\left(\theta,\eta,t\right)+\epsilon f_1\left(\theta,\eta,t\right)+... .
\end{equation}
Grouping the terms in like powers of $\epsilon$ gives  leading and higher order perturbation  equations; we only consider the first two orders here which is sufficient for our purpose. The $\mathcal{O}\left(\frac{1}{\epsilon}\right)$ equation is
\begin{equation}
\label{blo2}
-\omega f_{0,\theta}+k f_{0}f_{0,\theta}+k^{2}\mathcal{H}\left(f_{0,\theta\theta}\right)=0;
\end{equation}
and at $\mathcal{O}(1)$ by the linear equation
\begin{equation}
\label{bsop}
\mathcal{L}_{H}f_1\equiv-\omega f_{1,\theta}+k\left(f_{0}f_{1}\right)_{\theta}+k^{2}\mathcal{H}\left(f_{1,\theta\theta}\right)=G
\end{equation}
where
\begin{equation}
\label{gg2}
G\equiv-\left[f_{0,t}+f_{0}f_{0,\eta}+\mathcal{H}
\left(k_{\eta}f_{0,\theta}+2kf_{0,\theta\eta}\right)+\frac{f_0}{t+t_0}\right].
\end{equation}
The solution of Eq.~(\ref{blo2}) is
\begin{equation}
\label{blos}
f_0\left(\theta,\eta,t\right)=\frac{4k^2}{\sqrt{A^2+4k^2}-A \cos(\theta-\theta_0)}+\beta
\end{equation}
where $\theta_0$ is constant.

Eq.~(\ref{blos}) is the periodic wave solution of the classical BO equation\cite{Ben67}. Here $A=\frac{1}{2}\left(f_{0,max}-f_{0,min}\right)$ is the amplitude of the wave (cf. \cite{Mat07}) and the phase velocity of the wave is given by
\begin{equation}
\label{pv}
V=\frac{1}{2}\sqrt{A^2+4k^2}+\beta.
\end{equation}
In Eq.~(\ref{blos}), $k$, $A$, $\beta$ and $V$
are functions of slow variables $\eta$ and $t$. We will obtain the modulation equations for the sBO equation in terms of the three variables $k$, $V$ and $\beta$. Note that $A$ can be written in terms of these variables from Eq.~(\ref{pv}).

When the solution ~(\ref{blos}) is used in ~(\ref{bsop}), seculer terms occur, i.e. terms that grow arbitrarily large with respect to $\theta$. let $w$ denote solutions of the adjoint problem to $\mathcal{L}_{H}u=0$, i.e.,
\begin{equation}
\label{adj3}
\mathcal{L}_{H}^{A}w=0,\,\,\,\,\,\mathcal{L}_{H}^{A}= \omega \partial_{\theta}-k f_{0}\partial_{\theta}-k^{2}\mathcal{H}\left(\partial_{\theta\theta}\right)
\end{equation}
where we used the anti-symmetry property of the Hilbert transform: $\langle\mathcal{H}u,v\rangle=\langle u,-\mathcal{H}v\rangle, \langle,\rangle$ being the standard inner product. In order to eliminate secular terms, we use the following relation that follows from (\ref{bsop})
\begin{equation}
\label{adj3}
\int_{0}^{2\pi}[w\mathcal{L}_{H}f_{1}-f_{1}\mathcal{L}_{H}^{A}w]d\theta=\int_{0}^{2\pi}wGd\theta.
\end{equation}
We put $w=1$ and $w=f_0$ into Eq.~(\ref{adj3}), enforce the periodicity of $f_0\left(\theta,\eta,t\right)$ in $\theta$ and obtain the secularity conditions respectively as
\begin{equation}
\label{biconl1}
\frac{\partial}{\partial t}\int_{0}^{2\pi}f_{0}d\theta+\frac{\partial}{\partial \eta}\left(\frac{1}{2}\int_{0}^{2\pi}f_{0}^{2}d\theta\right)
+\frac{1}{t+t_0}\int_{0}^{2\pi}f_{0}d\theta=0,
\end{equation}
\begin{equation}
\label{biconl2}
\frac{\partial}{\partial t}\int_{0}^{2\pi}f_{0}^{2}d\theta+\frac{\partial}{\partial \eta}\left(\frac{2}{3}\int_{0}^{2\pi}f_{0}^{3}d\theta\right)+2\int_{0}^{2\pi}f_0\mathcal{H}
\left(k_{\eta}f_{0,\theta}+2kf_{0,\theta\eta}\right)d\theta+\frac{2}{t+t_0}\int_{0}^{2\pi}f_{0}^{2}d\theta=0.
\end{equation}
Computing the integrals in Eqs.~(\ref{biconl1}) and (\ref{biconl2}) with the properties of the Hilbert transform \cite{Ono75} and using the definition (\ref{pv}) we can obtain the following conservation laws
\begin{equation}
\label{bcclw2}
\beta_t+\beta\beta_{\eta}+\frac{2k+\beta}{t+t_0}=0
\end{equation}
and
\begin{equation}
\label{bcclw3}
V_t+VV_{\eta}+kk_{\eta}+\frac{2V-\beta}{t+t_0}=0.
\end{equation}
Equations~(\ref{comp}), (\ref{bcclw2}) and (\ref{bcclw3}) are the three conservation laws for the three variables $k$, $V$ and $\beta$.

We transform these conservation laws by using the following Riemann type variables $a,b,c$ \cite{Mat07} to simplify the conservation laws:
\begin{equation}
\label{brv}
k=b-a,\,\,\,\,\,\,\,\,V=b+a,\,\,\,\,\,\,\beta=2c
\end{equation}
and write the leading order solution $f_0$ in terms of  $a,b,c$
\begin{equation}
\label{bsoll}
f_0\left(\theta,\eta,t\right)=\frac{2\left(b-a\right)^2}{\left(b+a-2c\right)-2\sqrt{\left(a-c\right)\left(b-c\right)}\cos(\theta-\theta_0)}+2c
\end{equation}
The rapid phase $\theta$ is determined by integrating (\ref{phase})
\begin{equation}
\label{iphase}
\theta\left(\eta,t\right)=\int_{-L}^{\eta}\frac{k(x^{'},t)}{\epsilon}dx^{'}-\int_{0}^{t}\frac{k(\eta,t^{'})V(\eta,t^{'})}{\epsilon}dt^{'}
\end{equation}
Note that there is a free costant $\theta_0$ in Eq. (\ref{bsoll}). We determine the value of $\theta_0$ by comparison with direct numerical solutions.

By the transformation (\ref{brv}), we obtain the quasilinear PDE system for Riemann variables $a$, $b$ and $c$ in the following form
\begin{equation}
\label{bdiag}
\begin{aligned}
a_t+2aa_{\eta}+\frac{a+b-c}{t+t_0}=0,\\
b_t+2bb_{\eta}+\frac{a+b-c}{t+t_0}=0,\\
c_t+2cc_{\eta}+\frac{b+c-a}{t+t_0}=0.
\end{aligned}
\end{equation}
We take initial values of the Riemann variables $a,b,c$ of the Whitham system (\ref{bdiag})  to be steplike
\begin{equation}
\label{bwic}
a(\eta,0) = \left\{
\begin{array}{lr}
 0, &  \eta \leq0;\\
 \frac{1}{2}, &   \eta >0,
 \end{array}
 \right.\,\,\,\,\,\,\,\,b(\eta,0)=\frac{1}{2},\,\,\,\,\,\,\,\,c(\eta,0)=0.
\end{equation}

In the absence of spherical terms, i.e. $t_0\rightarrow\infty$, Eq. (\ref{bdiag}) reduces to a diagonal system that agrees with the Whitham system for the BO equation \cite{Mat98}. Corresponding to the initial condition (\ref{bwic}), the Whitham system (\ref{bdiag})  for the classical BO equation admits an exact rarefaction wave solution  in terms of the self-similar variable $\xi=\eta/t$. This solution is the dispersive regularization for the initial data  (\ref{bwic}) and  it is $a=a(\xi)= \xi/2$, $b=1/2$ and $c=0$.

But, the Whitham system (\ref{bdiag}) for sBO is not diagonal and this non-diagonal property prevents to get an analytical solution of the general Whitham system (\ref{bdiag}).  Therefore we follow the numerical approach developed for the sBO case in order to understand the structure of DSWs in the sBO equation.

\subsection{Comparison between numerical solutions}

Our aim is to investigate the sBO equation by using numerical methods to solve the general Whitham system (\ref{bdiag}). We compare the results of the Whitham system to direct numerical simulations of the 1+1 sBO equation and get very good agreement between them. We state that Whitham modulation theory
provides a good approximation of DSWs in the sBO equation. The benefit of studying with the Whitham system is about the order of coefficients. The Whitham system gives the structure of the DSWs in terms of $\mathcal{O}(1)$ coefficients. However, for direct numerical simulations one has small coefficients (due to $\epsilon \ll 1$) which in turn requires more complexity to solve. Also, direct numerical computation needs longer computing times.

First, we obtain numerical solutions of both reduced the Whitham system for the BO eq. and the non-diagonal Whitham system (\ref{bdiag}) for the sBO eq. The initial values of Riemann variables are given by Eq. (\ref{bwic}) (see Fig.~\ref{fig1}).

Before numerical computations we need to get the necessary boundary conditions for (\ref{bdiag}). The boundary conditions remain constant at both ends of the domain for the BO equation and associated Whitham system. However, the boundary conditions are functions of time for both the sBO equation and associated Whitham system (\ref{bdiag}). The boundary conditions  for the sBO equation are the same as in Eq.~(\ref{bc2}). We get the boundary conditions for the Whitham system (\ref{bdiag}) by solving analytically the reduced ODE system obtained from Eq. (\ref{bdiag}) by neglecting the spatial variable $\eta$. This ODE system is solved with the initial conditions (\ref{bwic}) at both left and right ends separately. The exact solutions for the boundary conditions on the left side are given by
\begin{equation}
\label{bbcl}
\begin{aligned}
a_{-}&=\frac{1}{2}\left[R(t)-1\right],\\
b_{-}&=\frac{R(t)}{2},\\
c_{-}&=\frac{1}{2}\left[R(t)-1\right]
\end{aligned}
\end{equation}
where $R(t)$ is given by (\ref{rr}). Similarly the boundary conditions on the right side (\ref{bwic}) are found to be
\begin{equation}
\label{bbcr}
\begin{aligned}
a_{+}=b_{+}=\frac{R^2(t)}{2},\,\,\,\,\,\,\,\,\,c_{+}=0.
\end{aligned}
\end{equation}
We see that all Riemann variables (except $c_+$) for the sBO equation at the boundaries decay in time.

To solve the Whitham system (\ref{bdiag}) with the boundary conditions (\ref{bbcl}) and (\ref{bbcr}), we use a first order hyperbolic PDE solver based on MATLAB$\circledR$ by Shampine \cite{Sha05} and choose a two-step variant of the Lax-Wendroff method with a nonlinear filter \cite{Eng89}. we use $N=2^{14}$ points for the spatial domain $[-30,30]$ with the time step being 0.9 times the spatial step in the numerical solutions of the Whitham system. The parameter in the sBO equation (\ref{sBO}) are taken to be as $t_0=10$. The results are given in Fig.~\ref{fig2} for both BO and sBO.

Next we solve the BO and sBO  equations (i.e.~Eq.~(\ref{sBO}) with $\tilde{c}=0$ and $\tilde{c} \neq 0$ respectively) numerically by using a modified version of the exponential time- differencing fourth-order Runge Kutta (ETDRK4) method. In the direct numerics, we use a regularization of the initial condition (\ref{ric}). See the section  \ref{Appe}  for the details of the version of ETDRK4 method that we use in computations. In this computation, we use $N=2^{15}$ spatial Fourier modes with the domain size $L=30$, and choose the time step to be $10^{-4}$. The regularization parameter in the initial condition (\ref{sic}) is chosen to be $\tilde{K}=10$, and the parameters in the sBO equation (\ref{sBO}) are taken to be $t_0=10$ and $\epsilon=10^{-3/2}$.  Direct numerical simulations of BO/sBO and solutions of the associated Whitham equations are compared in Fig.~\ref{fig3} at $t=7.5$. In the reconstructions from the Whitham equations, we fix the arbitrary constant phase $\theta_0$ in Eq.~(\ref{bsoll}) by adjusting the Whitham reconstruction to agree with those from direct numerical simulations. Details of this adjusting procedure are given below.

From the Fig.~\ref{fig2}, it should be noted that $b$ component of the solution of the Whitham system presents a small shock- like front in front of the DSW in the sBO case. This shock- like solution can be  regularized by adding higher order terms to the Whitham system \cite{Abl70}. Also, direct numerical simulations show that this behaviour is not important.

The DSW solutions can be generated for both BO and sBO at any time from the Riemann variables by using Eqs. (\ref{bsoll}) and (\ref{iphase}). These solutions are plotted and compared with direct numerical simulations of BO/sBO at the spesific time value $t=7.5$ in Fig.~\ref{fig3}a and Fig.~\ref{fig3}b. In Fig.~\ref{fig3}, the arbitrary phase $\theta_0$ in (\ref{bsoll}) is chosen to agree with the direct numerical simulations. For this adjustment, first we compute the average of the leading hump which as the largest amplitude and the
the trailing edge. Then $\theta_0$ is chosen as the center ('middle') of nearest wave  determined from asymptotic solution agrees with that of the corresponding hump determined from the direct numerical simulations. The average is approximately (3.387+1)/2=2.194 for the BO case and the asymptotic solution has a hump in the center region with an amplitude value of 2.127. For the sBO case (1.5426+ 0.5714)/2=1.0570 with a hump in the 'middle' region with an amplitude value of 1.039.

From the Fig.~\ref{fig3}, it is observed that the asymptotic approximation to the wave number and amplitude are in very good agreement with the direct numerics. Also, the humps in the 'middle' region of DSWs are captured well for both BO and sBO cases. However, it has to be reported that some phase deviations are noticed at the leading edge under enlargement of the figures. This phenomena is explained by Whitham Theory. According to theory, an order epsilon change in the phase $\theta$ causes an order one deviation in the location of humps at the leading edge. For the KdV eq., it was predicted in the work \cite{Cla10} by using integrable theory. Similar investigation can be performable for BO/sBO equations as a future study.

We also provide space-time plots of the direct numerical solutions of BO and sBO eqs in Fig.~\ref{fig4} to emphasize a difference between DSW solutions of these eqs. In Fig.~\ref{fig4}, it is observed that unlike the BO eq., the DSW humps in sBO eq. move to the right and spread apart. The spreading behaviour of DSW humps of sBO is also observed in an animation \cite{Mov1}. In the animation \cite{Mov1}, the propagation of DSWs in both BO and sBO eqs. between $t=0$ and $t=10$.

We report that the average speed of the leading hump of the DSW for BO eq. is approximately $V_{avg}=0.831$ at $t=7.5$. This speed is almost equal to the phase speed of the algebraic solitary wave solution of the BO equation \cite{Ono75} with an amplitude $4V_{avg}=3.34$. This solitary wave is approximately represented by
\begin{equation}
\label{solBO}
f(\eta,t)=\frac{4V}{1+\left[\frac{V\left(\eta-Vt\right)}{\epsilon}\right]^2}.
\end{equation}

However, the average speed of the leading hump of the DSW at $t=7.5$ for the sBO eq. is approximately $V_{avg}=0.4212$. Since the amplitude of the leading edge decreases in time, this speed is significantly smaller than the speed in the BO case. The  trailing edge of the DSW for the sBO eq. looks similar to that of the BO equation but its amplitude also decreases in time (see Fig.~\ref{fig3}).

The solution of the $1+1$ dimensional spherical BO equation coincides with the solution of the $3+1$ dimensional 3DBO only at $y^2+z^2=0$ (i.e. $y=z=0$). To construct the solution of the 3DBO Eq. at any circular cross section different from  $y^2+z^2=0$, the solution of the $1+1$ dimensional spherical BO Eq. must be shifted by a term which can be determined by the solution of the FS equation (\ref{center}) for the chosen circular cross section $y^2+z^2=r^2$ and time t. The shifting is performed by $x=\eta-\frac{r^2}{t^{*}+t_0}$ for a fixed time value $t^{*}$. Here $r$ is a fixed radius of the chosen cross section. By this shifting procedure, the evolution of DSWs on any cross section of the initial paraboloid in the 3DBO equation can be constructed for any time value $t$.

\section{Conclusion}

In this study, Dispersive shock waves in the 3DBO equation is considered using step like initial data along a paraboloid front. We use a similarity reduction method to reduce
$3+1$ dimensional BO equation to the $1+1$ dimensional spherical BO (sBO) equation. By using Whitham modulation theory, we obtain modulation equations and write these equations in terms of appropriate Riemmann type variables. We solve the resulting Whitham modulation equations numerically and compare these results with direct numerical solutions of the sBO equation. In this comparison, a good agreement is found between these numerics except an insignificant phase and a small discontinuity in front of the DSW. The discontinuity can be investigated by considering higher order terms; but this is outside the scope of this study.

According to our knowledge, this study is a first attempt to understand the structure of DSWs in $3+1$ dimensional systems. The method used in this work can be used to investigate the DSWs in other (3+1) dimensional equations. But the method introduced here works only for a special choice of initial front, e.g. a paraboloid front. This restriction is related to symmetry reductions of the corresponding $3+1$ dimensional equations. However, the reduction method used in this work can be applied to some $3+1$ dimensional equations such as $3+1$ dimensional Kadomtsev- Petviashvili (KP) equation and modified KP equation since these equations admit suitable symmetry reductions for choice of paraboloid front. We address these investigations for near future studies.

Also, DSWs in $3+1$ dimensional systems can be considered for more general class of initial conditions. To do that the method introduced in \cite{Abl17, Abl18} for Whitham modulation theory on $2+1$ dimensional systems must be generalized for Whitham modulation theory on $3+1$ dimensional systems. This prospective development of the theory will be topic of future studies, too.

\section{Acknowledgements}
This research was supported by the Istanbul Technical University Office of Scientific Research Projects (ITU BAPSIS), under grant TGA-2018-41318. We thank D.E. Baldwin for MATLAB codes of version of the ETDRK4 method that we use in the study.

\newpage

\bibliographystyle{elsarticle-num}

\begin{thebibliography}{9}

\bibitem{Ligh78} J. Lighthill, Waves in Fluids, \emph{Cambridge University Press}, Cambridge, UK, 1978.

\bibitem {Smy88} N.F. Smyth, P.E. Holloway, Hydraulic Jump and Undular Bore Formation on a Shelf Break, \emph{J. Phys. Oceanogr}. 18 (1988) 947-962.

\bibitem {Tay70} R.J. Taylor, D.R. Baker, H. Ikezi, Observation of Collisionless Electrostatic Shocks, \emph{Phys. Rev. Lett}. 24(5) (1970) 206-209.

\bibitem {Wan07} W. Wan, S. Jia, J.W. Fleischer, Dispersive Superfluid-like Shock Waves in Nonlinear Optics, \emph{Nat. Phys}. 3 (2007) 46-51.

\bibitem {Con09} C. Conti, A. Fratalocchi, M. Peccianti, G. Ruocco, S. Trillo, Observation of a gradient catastrophe generating solitons,
          \emph{Phys. Rev. Lett}. 102 (2009) 083902.

\bibitem {Fat14} J. Fatome, C. Finot, G. Millot, A. Armaroli, S. Trillo, Observation of Optical Undular Bores in Multiple Four-Wave Mixing, \emph{Phys. Rev. X}.
          4  (2014) 021022.
          
\bibitem {Hoe06} M.A. Hoefer, M.J. Ablowitz, I. Coddington, E.A. Cornell, P. Engels, V. Schweikhard, Dispersive and Classical Shock Waves in Bose-Einstein
     Condensates and Gas Dynamics, \emph{Phys. Rev. A}. 74 (2006) 023623.

\bibitem{Hoe08} M.A. Hoefer, M.J. Ablowitz and P. Engels,  Piston dispersive shock wave problem, {\it Phys. Rev. Lett.}, 100, (2008) 084504.          

\bibitem {Whi65} G.B. Whitham, Non-linear Dispersive Waves, \emph{Proc. R. Soc. A}. 283 (1965) 238-261.

\bibitem {Gru74} A.V. Gurevich, L.P. Pitaevskii, Nonstationary Structure of a collisionless Shock Wave,\emph{ Sov. Phys. JETP-USSR}. 38(2) (1974) 291-297.

\bibitem{Lax83} P. Lax, C. Levermore, The Small Dispersion Limit of The Korteweg-De Vries Equation 1, \emph{Commun. Pure Appl. Math}. 36(3) (1983) 253-290.

\bibitem{Dris76} C. Driscoll, T. O'Neil, Modulational instability of cnoidal wave solutions of the modified Korteweg-de Vries equation, \emph{J. Math. Phys.} 17(7) (1976) 1196-1200.

\bibitem {Gru87} A.V. Gurevich, L.P. Pitaevskii, Averaged Description of Waves in the Korteweg- de Vries-Burgers Equation, \emph{Zh. Eksp. Teor. Fiz}. 93 (1987)
          871-880.

\bibitem {Mat98} Y. Matsuno, Nonlinear Modulation of Periodic Waves in the Small Dispersion Limit of the Benjamin-Ono Equation, \emph{Phys. Rev. E}.
              58(6) (1998) 7934-7939.

\bibitem {Mat07} Y. Matsuno, V.S. Shchesnovich, A.M. Kamchatnov, R.A. Kraenkel, Whitham Method For the Benjamin-Ono-Burgers Equation And Dispersive Shocks,
               \emph{Phys. Rev. E}. 75 (2007) 016307.

\bibitem {Kamc12} A.M. Kamchatnov, Y.H. Kuo, T.C. Lin, T.L. Horng, S.C. Gou, R. Clift, G.A. El, R.H.J. Grimshaw, Undular Bore theory for the Gardner equation, \emph{Phys. Rev. E.} 86 (2012) 036605.

\bibitem {Hoe16} G. El, M. Hoefer, Dispersive Shock Waves and Modulation Theory, \emph{Physica D.} 333 (2016) 11-65.

\bibitem{EL09a} G.A. El, A.M. Kamchatnov, V.V. Khodorovskii, E.S. Annibale, A. Gammal, Two-dimesional Supersonic Nonlinear Schrödinger Flow Past an Extended Obstacle,
        \emph{Phys. Rev. E.} 80 (2009) 046317.

\bibitem{Hoe09a} M.A. Hoefer, B. Ilan, Theory of Two-dimesional Oblique Dispersive Shock Waves in Supersonic Flow a Superfluid, \emph{Phys. Rev. E.} 80 (2009) 061601(R).

\bibitem{KP70} B.B. Kadomtsev, V.I. Petviashvili, On the Stability of Solitary Waves in Weakly Dispersing Media,
         \emph{Sov. Phys. Dokl}. 15 (1970) 539.

\bibitem{Abl80} M.J. Ablowitz, H. Segur, Long Internal Waves in Fluids of Great Depth, \emph{Stud. Appl. Math}. 62 (1980) 249-262.


\bibitem{Dem16} M.J. Ablowitz, A. Demirci, Y.P. Ma, Dispersive Shock Waves in the Kadomtsev- Petviashvili and Two Dimensional Benjamin- Ono Equations, \emph{Physica D.} 333 (2016) 84-98.


\bibitem {Abl17}  M.J. Ablowitz, G. Biondini, Q. Wang, Whitham Modulation Theory For the Two- Dimensional Benjamin- Ono Equation, \emph{Phys. Rev. E.} 96 (2017) 032225.

\bibitem {Abl18} M.J. Ablowitz, G. Biondini, I. Rumanov, Whitham Modulation Theory For (2+1)- Dimensional Equations of Kadomtsev- Petviashvili Type, \emph{J. Phys. A: Math. Theor.} 51 (2018) 215501.

\bibitem{El13a} G.A. El, Hydrodynamic Type Systems and Their Integrability, \emph{http://www.researchgate.net/publication/258997674},(2013).

\bibitem {Luk66} J.C. Luke, A Perturbation Method for Nonlinear Dispersive Wave Problems, \emph{Proc. R. Soc. A}. 292 (1966) 403-412.

\bibitem {Abl70} M.J. Ablowitz, D.J. Benney, Evolution of Multi-Phase Modes For Nonlinear Dispersive Waves, \emph{Stud. Appl. Math}. 49(3) (1970) 225.

\bibitem{Gun18} J.M. Conde, F. Gungor, Analysis of the Symmetry Group and Exact Solutions of the Dispersionless KP Equation in n+1 Dimensions, \emph{J. Math. Phys.} 59 (2018) 111501.

\bibitem {Ben67} T.B. Benjamin, Internal Waves of Permanent Form in Fluids of Great Depth, J. Fluid Mech. 29(3) (1967) 559-592.

\bibitem {Ono75} H. Ono, Algebraic Solitary Waves in Stratified Fluids, \emph{J. Phys. Soc. Jpn}. 39 (1975) 1082.

\bibitem{Kle11a} C. Klein, K. Roidot, Fourth Order Time-Stepping for Kadomtsev-Petviashvili and Davey-Stewartson Equations, \emph{SIAM J. Sci. Comput.} 33(6) (2011) 33333356.

\bibitem {Abl13} M.J. Ablowitz, D.E. Baldwin, Dispersive Shock Wave Interactions and Asymptotics, \emph{Phy. Rev. E}. 87 (2013) 022906.

\bibitem {Sha05} L.F. Shampine, Solving Hyperbolic PDEs in MATLAB, \emph{Appl. Numer. Anal. Comput. Math}. 2(3) (2005) 346-358.

\bibitem {Eng89} B. Engquist, P. Lötstedt, B. Sjögreen, Nonlinear Fillters For Efficient Shock Computation, \emph{Math. Comp}. 52 (1989) 509-537.

\bibitem {Cla10}T. Claeys, T. Grava, Solitonic Asymptotics For the Korteweg-de Vries Equation in the Small Dispersion Limit, \emph{SIAM J. Math. Anal}. 42 (2010) 2132-2154.

\bibitem {Mov1} Dispersive Shock Wave Propagation in Benjamin- Ono and Spherical Benjamin- Ono Equations Between $t=0$ and $t=10$, \href{https://youtu.be/K8FOtgYcKYc}{$youtu.be/K8FOtgYcKYc$}.

\bibitem {Abl09} M.J. Ablowitz, D.E. Baldwin, M.A. Hoefer, Soliton Generation and Multiple Phases in Dispersive Shock and Rarefaction Wave Interaction,
       \emph{Phys. Rev. E}. 80(1) (2009) 016603.

\bibitem {Cox02} S.M. Cox, P.C. Matthews, Exponential Time Differencing For Stiff Systems, \emph{J. Comput. Phys}. 176 (2002) 430-455.

\bibitem {Kas05} A.K. Kassam, L.N. Trefethen, Fourth-order time-stepping For Stiff PDEs, \emph{SIAM J. Sci. Comput}. 26(4) (2005) 1214-1233.




\end{thebibliography}
\biboptions{compress}

\section{Appendix}
\label{Appe}

We use a numerical method for the direct numerical simulations which can be used for problems with fixed boundary conditions. But, left boundary condition for sBO is function of $t$. For this reason, we first transform (\ref{sBO}) by
\begin{equation}
\label{tr1}
f=R(t)\phi
\end{equation}
to the following variable coefficient BO (vBO) equation
\begin{equation}
\label{vkd}
\phi_{t}+R(t)\phi\phi_{\eta}+\epsilon\mathcal{H}\left(\phi_{\eta\eta}\right)=0.
\end{equation}
Here $R(t)=\frac{t_0}{t+t_0}$. Equation~(\ref{vkd}) has the left boundary condition fixed at $\phi_{-}=1$, while the right boundary condition $\phi_{+}=0$ stays the same as in the original sBO equation. In order to solve {Eq.~(\ref{vkd}) numerically (see also \cite{Abl09, Cox02, Kas05}) we differentiate with respect to $\eta$ and define $\phi_{\eta}=z$ to get
\begin{equation}
\label{zeq}
z_t+R(t)\left(z\phi\right)_{\eta}+\epsilon\mathcal{H}\left(z_{\eta\eta}\right)=0.
\end{equation}
Transforming to Fourier space gives
\begin{equation}
\label{fzeq}
 \widehat{z_t}= \textbf{L}\widehat{z}+R(t)\textbf{N}\left(\widehat{z},t\right)
\end{equation}
where $\widehat{z}=\mathcal{F}(z)$ is the Fourier transform of $z$, $\textbf{L}\widehat{z}\equiv i\epsilon sgn(k) k^2 \widehat{z}$ and
\begin{equation}
\label{nop}
\textbf{N}\left(\widehat{z},t\right)=-ik\mathcal{F}\left\{\left[\phi_{-}+\int_{-L}^{\eta}\mathcal{F}^{-1}\left(\widehat{z}\right)d\eta^{'}\right]
\mathcal{F}^{-1}\left(\widehat{z}\right)\right\}.
\end{equation}
where $L$ is a large positive constant. The only difference from the classical BO case is that for sBO the nonlinear term $\textbf{N}$ has a time dependent coefficient.
To solve the above ODE system in Fourier space we use a modified version of the exponential-time-differencing fourth-order Runge-Kutta (ETDRK4) method \cite{Cox02, Kas05}. For the required spectral accuracy of the ETDRK4 method, the initial condition for $z$ must be smooth and periodic. However, the step initial condition (\ref{ric}) for $u$ or equivalently $f$ leads to $z(\eta,0)=-\delta(\eta)$, where $\delta$ represents the Dirac delta function. Therefore we regularize this initial condition with the analytic function \cite{Abl09}
\begin{equation}
\label{sic}
z\left(\eta,0\right)=-\frac{\tilde{K}}{2}\textrm{sech}^2\left(\tilde{K}\eta\right),
\end{equation}
where $\tilde{K}>0$ is large. Thus Eq.~(\ref{zeq}) can be solved numerically via Eqs.~(\ref{fzeq}-\ref{nop}) on a finite spatial domain $[-L,L]$, where $\mathcal{F}$ represents the discrete Fourier transform.

\newpage

\section*{FIGURES}

\begin{figure}[ht]
\centering
\includegraphics[width=0.48\textwidth]{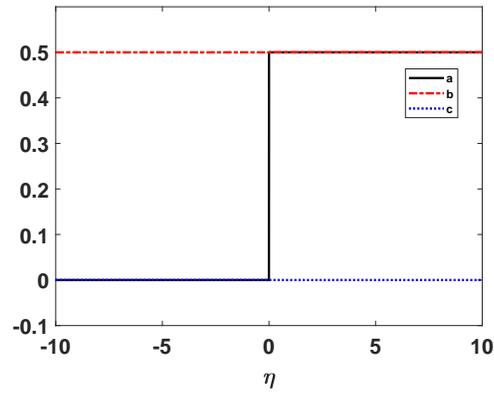}
\caption{\small Initial values (\ref{bwic}) for Riemann variables $a$, $b$ and $c$.}
\label{fig1}
\end{figure}

\begin{figure}[ht]
\centering
\subfigure[]{
\includegraphics[width=0.48\textwidth]{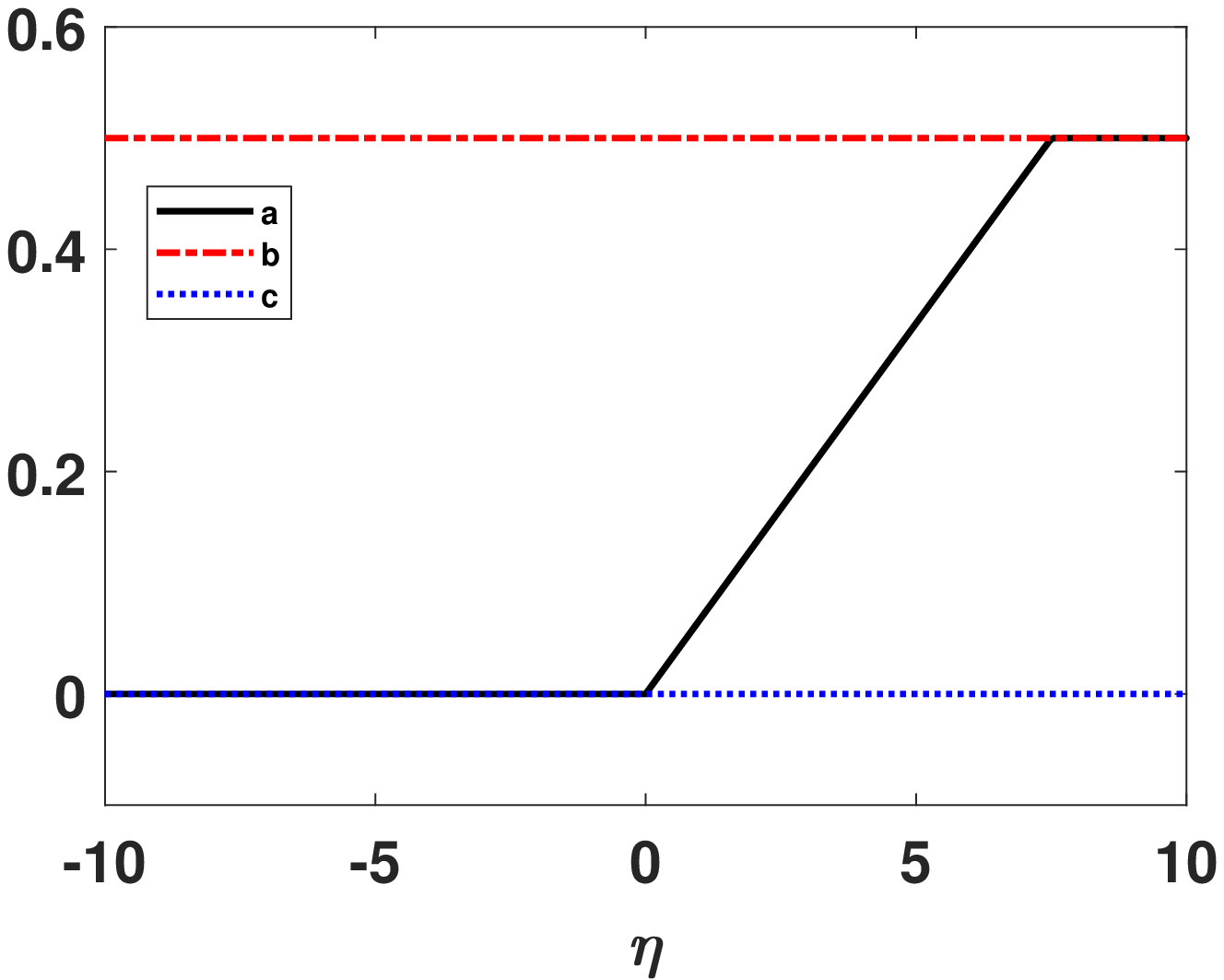}}
\subfigure[]{
\includegraphics[width=0.48\textwidth]{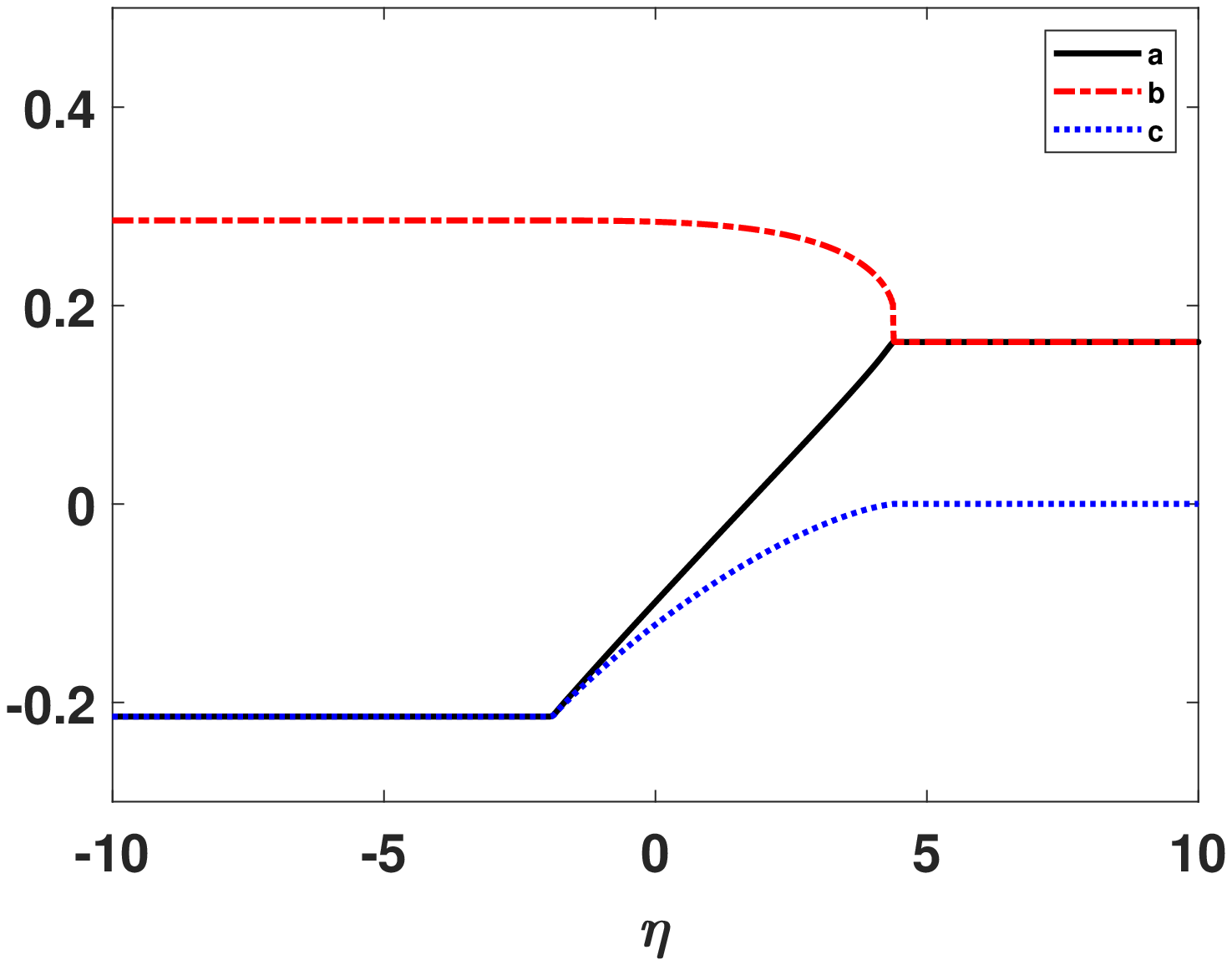}}
\caption{\small Riemann variables at t=7.5 which are found by numerical solutions of reduced Whitham system for BO eq. and exact Whitham system (\ref{bdiag}) for sBO eqs. (a) for BO eq., (b) for sBO eq. Here, we take $t_0=10$.}
\label{fig2}
\end{figure}

\begin{figure}[ht]
\centering
\subfigure[]{
\includegraphics[width=0.70\textwidth]{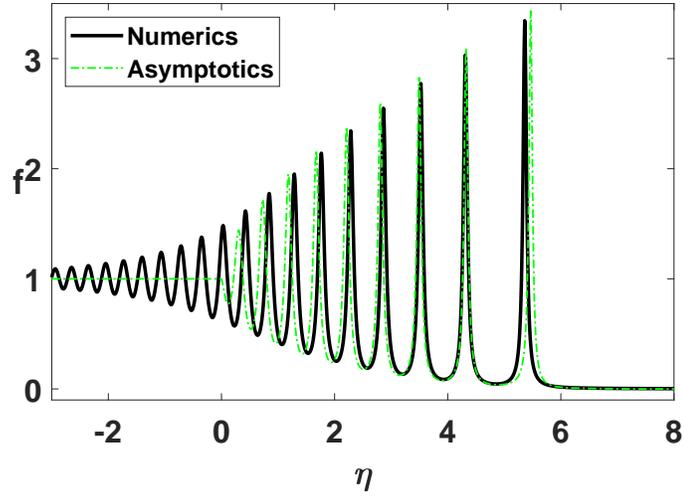}}
\subfigure[]{
\includegraphics[width=0.70\textwidth]{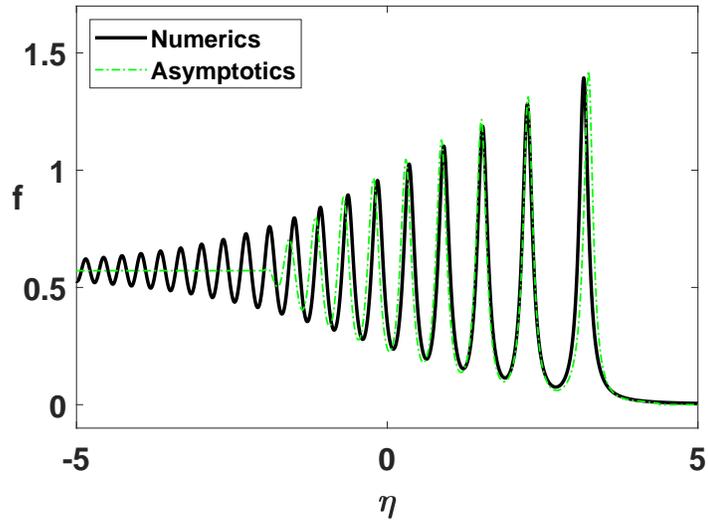}}
\caption{\small Numerical and asymptotic solutions of BO and sBO eqs. at t=7.5 with the initial data (\ref{ric}). (a)for BO eq., (b) for sBO eq. Here, we take $t_0=10$ and $\epsilon=10^{-3/2}$.}
\label{fig3}
\end{figure}

\begin{figure}[ht]
\centering
\begin{tabular}{cc}
\includegraphics[width=0.48\textwidth]{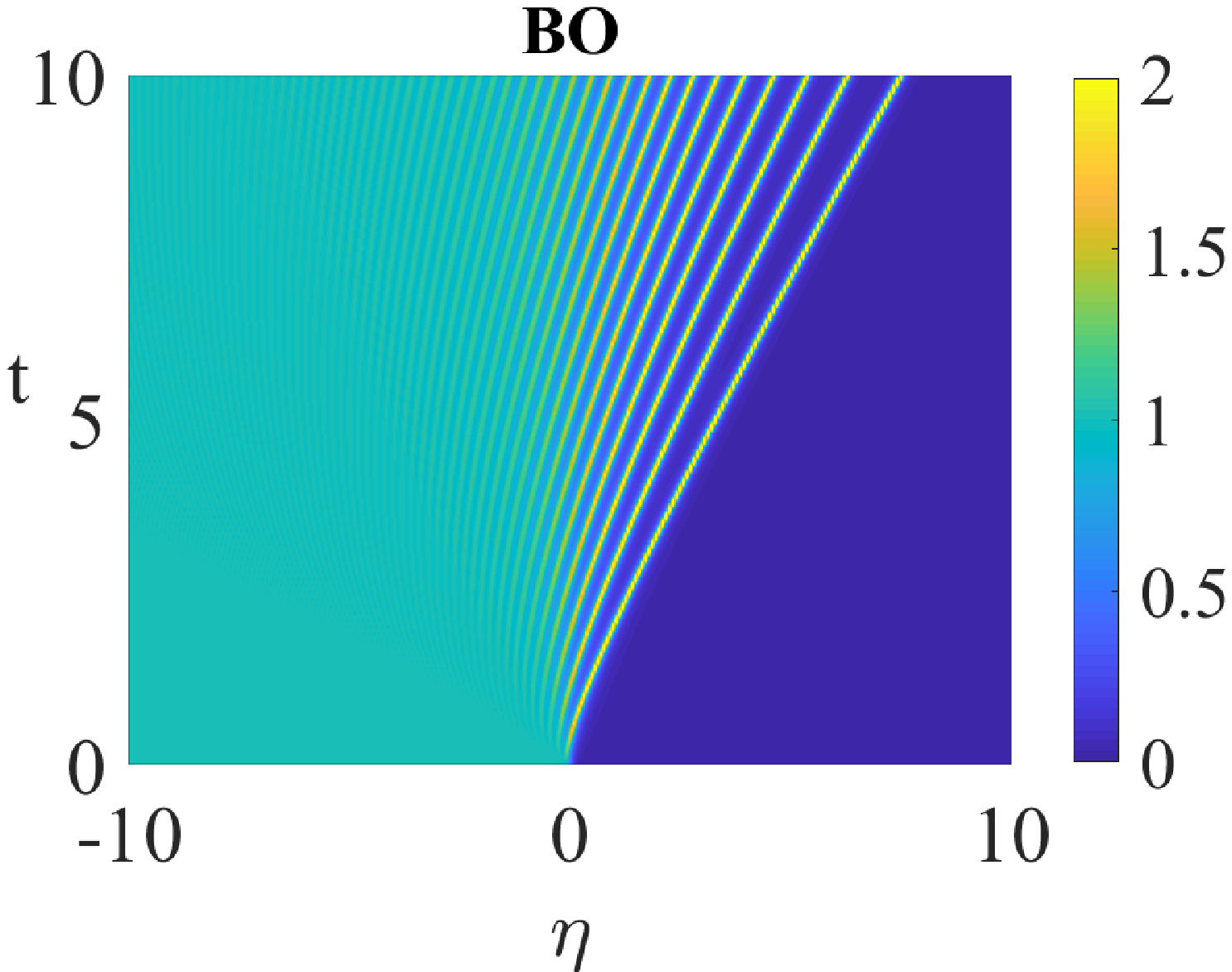}&
\includegraphics[width=0.48\textwidth]{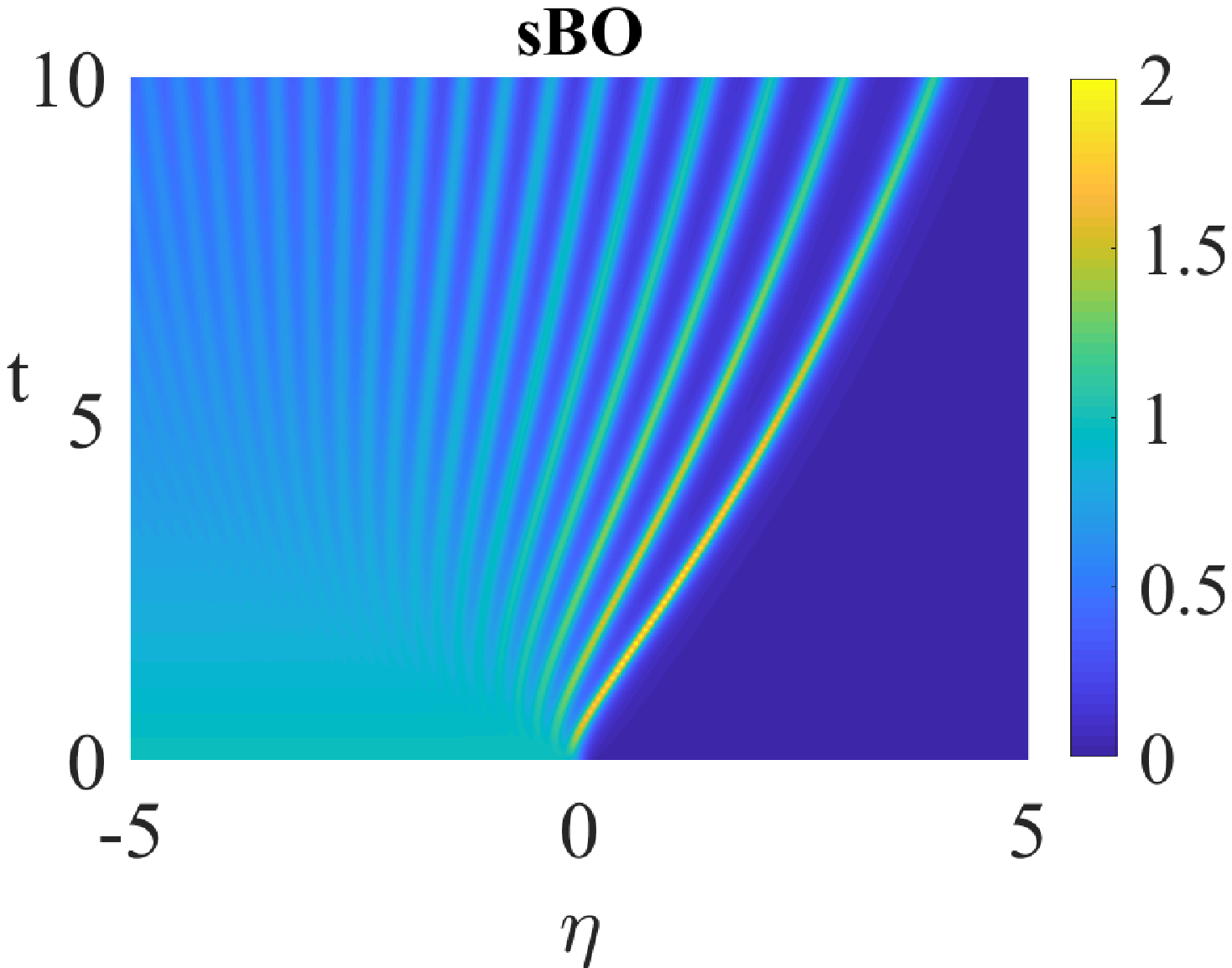}\\
(a) & (b)
\end{tabular}
\caption{\small Space-time plot of the direct numerical solutions between $t=0$ and $t=20$ (a)for BO eq., (b) for sBO eq. Here, we take $t_0=10$ and $\epsilon=10^{-3/2}$.}
\label{fig4}
\end{figure}

\end{document}